\documentclass[11pt, oneside]{article}   	
\usepackage{geometry}                		
\usepackage{graphicx}				
\usepackage{amssymb}
\usepackage{amsmath}
\usepackage{subfig}
\usepackage{caption}
\usepackage{comment}
\usepackage{amsthm,verbatim,amsfonts,amscd,float,physics}
\usepackage{appendix}

\usepackage[utf8]{inputenc}
\usepackage{xcolor}
\numberwithin{equation}{section}
\begin{document}
\title{Effective Matrix Model for Gauge Theories at Finite Temperature and Density using Quantum Computing }
\author{Yuan Feng$^{(1)}$,  Michael McGuigan$^{(2)}$ \\
(1) Pasadena City College,  (2) Brookhaven National Laboratory}
\date{}
\maketitle
\begin{abstract}
We study the effective matrix model for for gauge fields and fermions on a quantum computer. We use the Variational Quantum Eigensolver (VQE) using IBM QISKit for the effective matrix model for $SU(2)$ and $SU(3)$ including fermions in the fundamental representation.  For $SU(2)$ we study the effects of finite temperature and nonzero chemical potential.  In all cases we find excellent agreement with the classical computation.
\end{abstract}
\newpage

\section{Introduction}

Effective matrix models for gauge field theory have been devised to capture many of the important features of gauge theory however without the computational complexity of the full theory. The effective matrix model method is a good match for current quantum computers as it does not take nearly as many qubits to represent the Hamiltonian of the effective matrix model as does a lattice Hamiltonians such as the those introduced by Kogut and Susskind. The effective matrix model has a broad range of applications including QCD \cite{KorthalsAltes:1994xx}\cite{Dumitru:2012fw}\cite{Kashiwa:2012wa},  high energy physics applications to Wilson line symmetry breaking, gauge-Higgs unification \cite{Shiraishi:1986wu}
\cite{Shiraishi:1989gt}
\cite{Shiraishi:1986wu}
\cite{Ho:1990xz}
\cite{Davis:2005qi}
\cite{Hosotani:1983xw}
\cite{Gursoy:2007np}
\cite{Itoyama:2020ifw}
\cite{Hamada:2015ria}
\cite{Ginsparg:1986wr}
\cite{Nair:1986zn}
\cite{Faraggi:2000pv}
\cite{McGuigan:2019gdb}\cite{Yamanaka:2019aeq}
as well condensed matter and nanoscience applications through the computation of the persistent current \cite{Duong}. All these applications of effective matrix models can be run efficiently on near term quantum computers. In  \cite{Miceli:2019snu} we studied the effective matrix model and quantum computing for $SU(2)$ gauge theories and in this paper we extend that work to finite temperature, finite density and $SU(3)$ gauge theory.

\section{ VQE for effective Matrix model for $SU(2)$ } 

Starting with the Lagrangian of gauge fields coupled to fermions as
\begin{equation}L =  - \frac{1}{4}{F_{\mu \nu }}{F^{\mu \nu }} + \bar \psi i\gamma^\mu D_\mu\psi \end{equation}
with:
\[{F_{\mu \nu }} = {\partial _\mu }{A_\nu } - {\partial _\nu }{A_\mu } + ig\left[ {{A_\mu },{A_\nu }} \right]\]
\begin{equation}{D_\mu }\psi  = \left( {{\partial _\mu } + ig{T^a}A_\mu ^a} \right)\psi \end{equation}
one can derive a one-loop appriximation for the Effective Matrix Model fpr $SU(2)$. In the effective Matrix model one has a compact direction say $x_3$ with the topology  of $S^1$ and uses an ansatz where the gauge field is diagonal so:
\begin{equation}A_3^a(t){\sigma ^a} = \left( {\begin{array}{*{20}{c}}
{\phi (t(}&0\\
0&{ - \phi (t)}
\end{array}} \right)\end{equation}
with $x_3 \rightarrow x_3 + 2\pi L$ where $L$ is the radius of the $S^1$. 
The potential for the one-loop approximation is the sum of two terms. One term is $V_g$ associated with the gauge bosons and another term $V_f$ associated with fermions. In terms of determinants of differential operators they are given by:
\[{V_g} = \frac{1}{2}2\ln \det ( - {D^2})\]
\begin{equation}{V_f} =  - \frac{1}{2}4\ln \det ( - {D^2})\end{equation}
The effective potential for $SU(2)$ with $N_F$ fermions in the one-loop approximation is given in \cite{Ho:1990xz}. 
\begin{equation}{V_{eff}}(\phi ) =  - \frac{2}{{{L^4}{\pi ^2}}}\sum\limits_{\ell  = 1}^\infty  {\frac{1}{{{\ell ^4}}}\left[ {2\cos (\ell \phi ) + 1} \right]}  + {N_F}\frac{4}{{{L^4}{\pi ^2}}}\sum\limits_{\ell  = 1}^\infty  {\frac{1}{{{\ell ^4}}}\left[ {2\cos (\ell \phi /2)} \right]} \end{equation}
The first term comes from the gauge bosons and the second from the fermions.
The potential is shown in figure 1. In this section we wish to calculate the ground state energy for  potentials of this type using the Variational Quantum Eigensolver and IBM QISKit. The effective Matrix Model Hamiltonian we consider for $SU(2)$ is:
\begin{equation}H = \frac{1}{2}p_\phi ^2 + {V_{eff}}(\phi )\end{equation}
The first step in quantum simulation is to represent the Hamiltonian in a finite Hilbert space representation which can be mapped to qubits. A convenient basis is the harmonic oscillator basis where the $p_\phi$ and $\phi$ operators can be represented as:
\begin{equation} 
 {\phi} = \frac{1}{\sqrt{2}}\begin{bmatrix}
 
   0 & {\sqrt 1 } & 0 &  \cdots  & 0  \\ 
   {\sqrt 1 } & 0 & {\sqrt 2 } &  \cdots  & 0  \\ 
   0 & {\sqrt 2 } &  \ddots  &  \ddots  & 0  \\ 
   0 & 0 &  \ddots  & 0 & {\sqrt {N-1} }  \\ 
   0 & 0 &  \cdots  & {\sqrt {N-1} } & 0  \\ 
\end{bmatrix}
  \end{equation}
while for the momentum operator we have:
\begin{equation}
 p_{\phi} = \frac{i}{\sqrt{2}}\begin{bmatrix}
 
   0 & -{\sqrt 1 } & 0 &  \cdots  & 0  \\ 
   {\sqrt 1 } & 0 & -{\sqrt 2 } &  \cdots  & 0  \\ 
   0 & {\sqrt 2 } &  \ddots  &  \ddots  & 0  \\ 
   0 & 0 &  \ddots  & 0 & -{\sqrt {N-1} }  \\ 
   0 & 0 &  \cdots  & {\sqrt {N-1} } & 0  \\ 
\end{bmatrix}
  \end{equation}

\begin{figure}
    \centering
    \begin{minipage}[b]{0.7\linewidth}
      \centering
      \includegraphics[width=\linewidth]{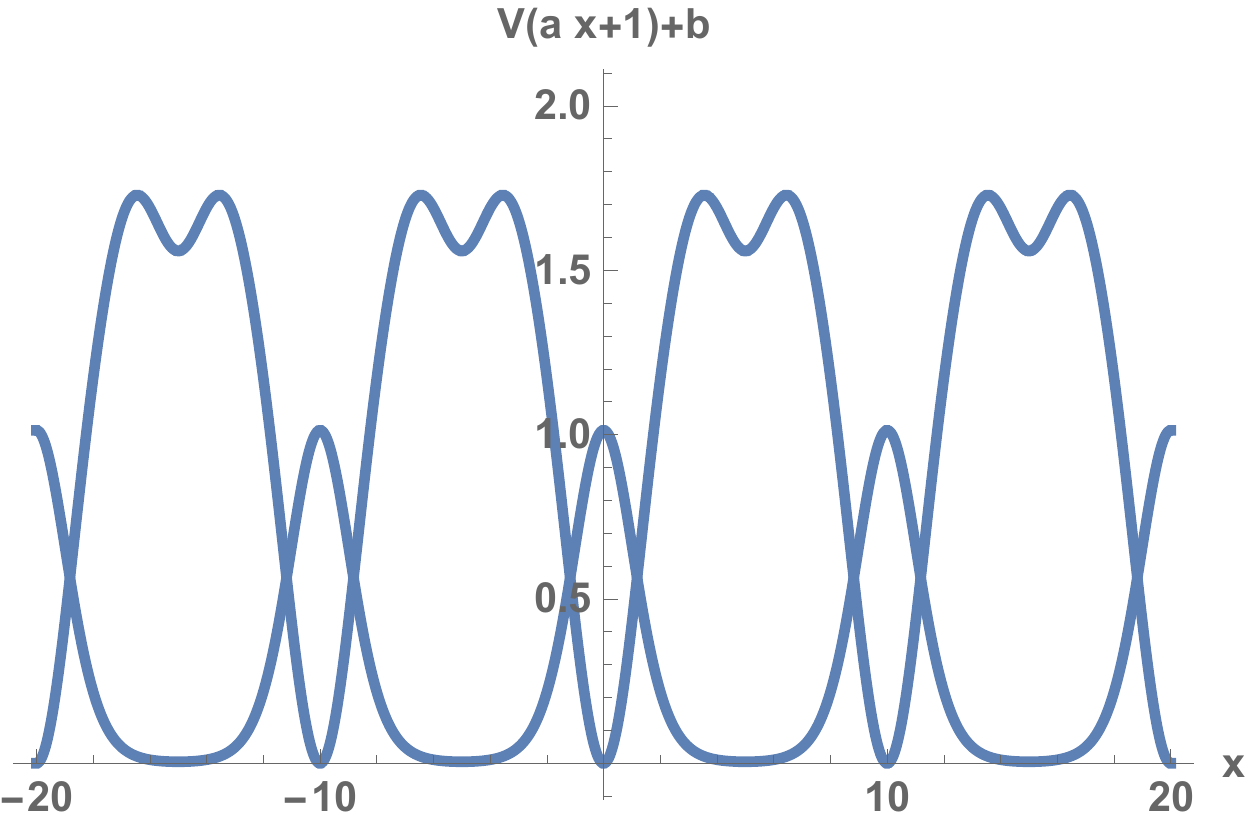}
    \end{minipage}
    \caption{Potential for $SU(2)$ Effective Matrix Model with one fermion in fundamental representation.The ground state wave function is localized in the valleys  of the potential.}
    \label{3-qubit_unit}
\end{figure}
\begin{figure}
    \centering
    \begin{minipage}[b]{0.7\linewidth}
      \centering
      \includegraphics[width=\linewidth]{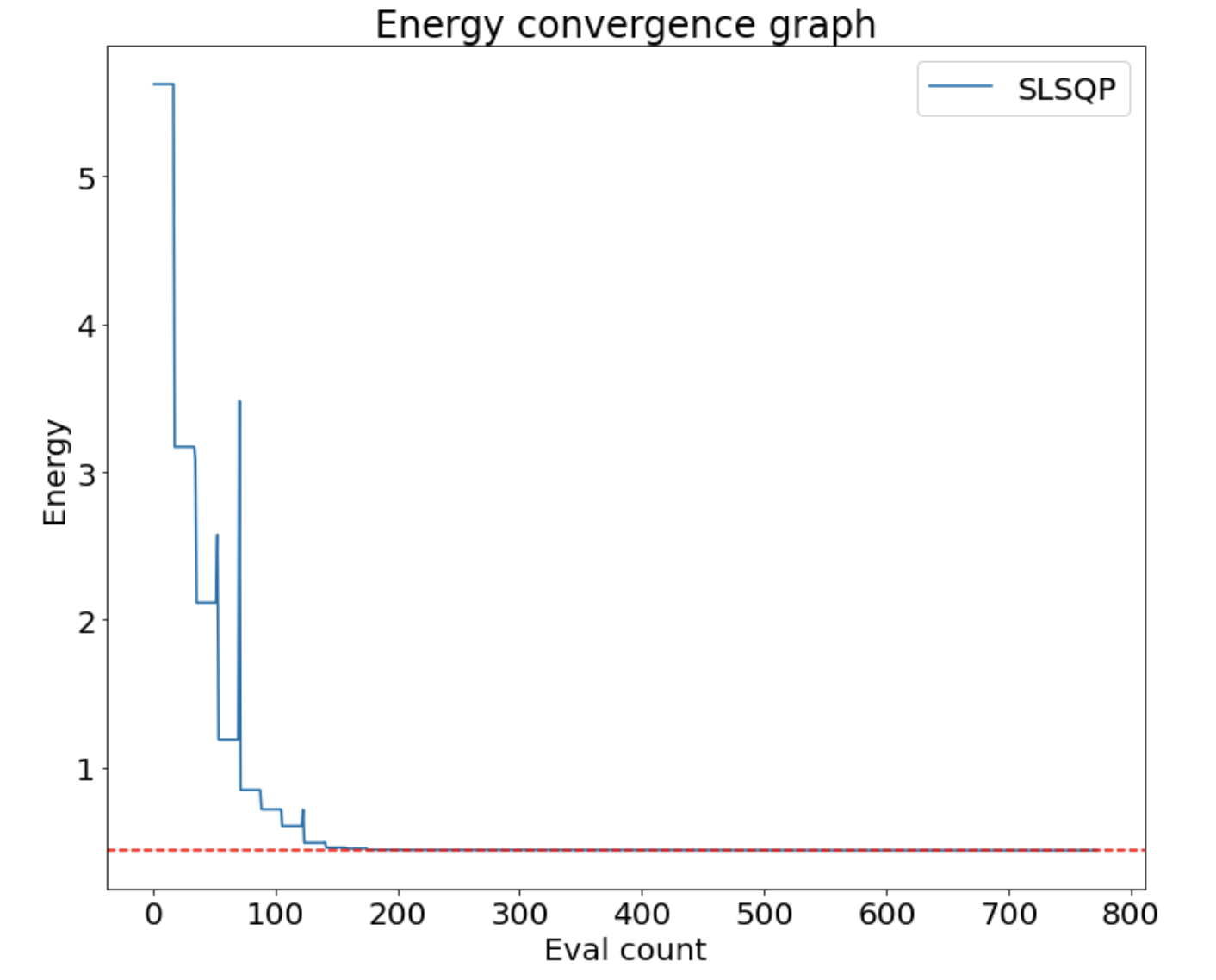}
    \end{minipage}
    \caption{VQE convergence plot for $SU(2)$ Effective Matrix Model with one fermion in fundamental representation.}
    \label{3-qubit_unit}
\end{figure}
   \begin{figure}
   \vspace{-4cm}
      \includegraphics[width=.7\linewidth, angle = 270 ]{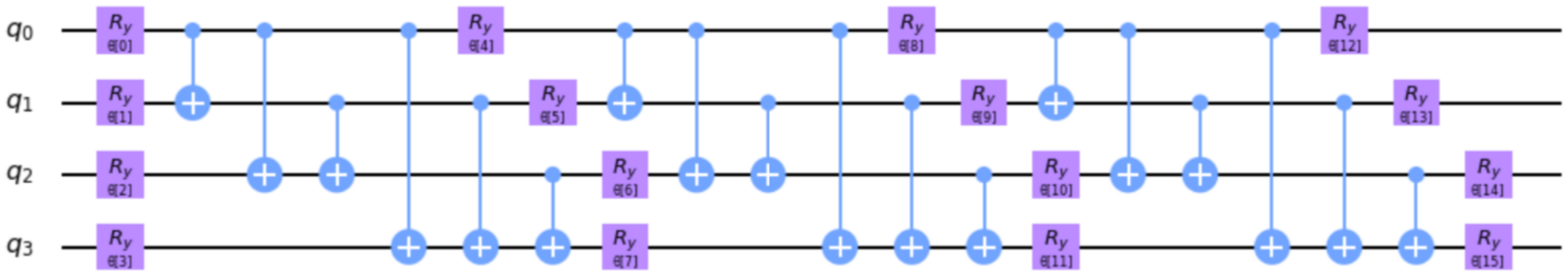}
        \vspace{-4cm}
      \centering
    \caption{Quantum circuit representing the variational ground state for the $SU(2)$ effective Matrix model with a fundamental fermion.}
    \label{3-qubit_unit}
\end{figure}
The Variational Quantum Eigensolver (VQE) is a hybrid classical-quantum algorithm based on the variational method of quantum mechanic to estimate the ground state energy and ground state wave function of a quantum Hamiltonian. By choosing a variational wave function $\psi$ one minimizes:
\begin{equation}{E_{{\mathop{\rm var}} }} = \frac{{\left\langle {\psi ({\theta _i})} \right|H\left| {\psi ({\theta _i})} \right\rangle }}{{\left\langle {\psi ({\theta _i})} \right|\left. {\psi ({\theta _i})} \right\rangle }}\end{equation}
The $\theta_i$ are angles which parameterize  the variational wave function and are represented on the quantum computer in terms of quantum gates. The Hamiltonian is represented in terms of qubits as a expansion of tensor products of two by two matrices given by the Pauli spin matrices plus the two by two identity matrix. One uses an optiimizer to find the minimum of $E_{var} $ over several iterations and over several runs to determine the least upper bound on the ground state energy. It is this least upper bound which is the result of the VQE that is its estimate to the ground state energy.
Using $16\times 16$ matrices for $\phi$ so the total The Hamiltonian is also a $16 \times 16$ matrix and can be represented by four qubits. Using the VQE and the SLSQP optimizer we find the results in table 1 which are  in excellent agreement with classical computation.
\begin{table}[ht]
\centering
\begin{tabular}{|l|l|l|l|l|}
\hline
Hamiltonian       & Qubits & Paulis & Exact Result & VQE Result \\ \hline
$SU(2)$ with  fermion in fundamental  & 4 & 71 & 0.4425673 & 0.4426310\\ 
 \hline

\end{tabular}
\caption{\label{tab:BasisCompare}  VQE results for effective Matrix model for $SU(2)$ gauge theory with a fermion in the fundamental representation and using the oscillator basis. The Hamiltonian were mapped to 4-qubit operators with 71 Pauli terms.  The quantum circuit for each simulation utilized an \(R_y\) variational form, with a fully entangled circuit of depth 3. The backend used was a statevector simulator. The Sequential Least SQuares Programming (SLSQP) optimizer was used, with a maximum of 600 iterations.}
\end{table}

\section{ VQE  for effective Matrix model for finite temperature } 

Finite temperature can be included by using the imaginary time formulation with periodic boundary conditions for bosons and antiperiodic boundary conditions for fermions in imaginary time.
For finite temperature the effective potential is given by \cite{Ho:1990xz}:
\begin{equation}{V_{eff}}(\phi ,T) = {V_{eff}}(\phi ) + {V_{eff}}(\phi ,T \ne 0)\end{equation}
where:
\begin{equation}{V_{eff}}(\phi ,T \ne 0) =  - \frac{2}{{{\pi ^2}}}\sum\limits_{m =  - \infty ,\ell  = 1}^\infty  {\frac{{\cos (2\ell \phi )}}{{{{\left[ {{L^2}{\ell ^2} + {\beta ^2}{m^2}} \right]}^2}}}}  + {N_F}\frac{4}{{{L^4}{\pi ^2}}}\sum\limits_{m =  - \infty ,\ell  = 1}^\infty  {\frac{{{{( - 1)}^m}\cos (\ell \phi )}}{{{{\left[ {{L^2}{\ell ^2} + {\beta ^2}{m^2}} \right]}^2}}}} \end{equation}
The potential is plotted in figure 4. For small $\beta$ this can be approximated by:
\begin{equation}{V_{eff}}(\phi ,T) \approx  - VL\frac{{{\pi ^2}}}{{90}}\left( {2 \times 3 + \frac{7}{8}4{N_F}} \right)\frac{1}{{{\beta ^4}}} - \frac{{2V}}{{{L^3}}}\frac{L}{\beta }\left[ {2\sum\limits_{\ell  = 1}^\infty  {\frac{{\cos (2\ell \phi )}}{{{\ell ^3}}} + \zeta (3)} } \right]\end{equation}
For a representative calculation we choose a temperature such that $\frac{V L}{L^3 \beta} = 1$. Then the exact ground state energy is $E_0(\beta) =.46617183 $. Using the VQE algorithm we find from table 2, $E_0(\beta) = 0.46617228 $ in excellent agreement with the classical computation.
\begin{figure}
    \centering
    \begin{minipage}[b]{0.7\linewidth}
      \centering
      \includegraphics[width=\linewidth]{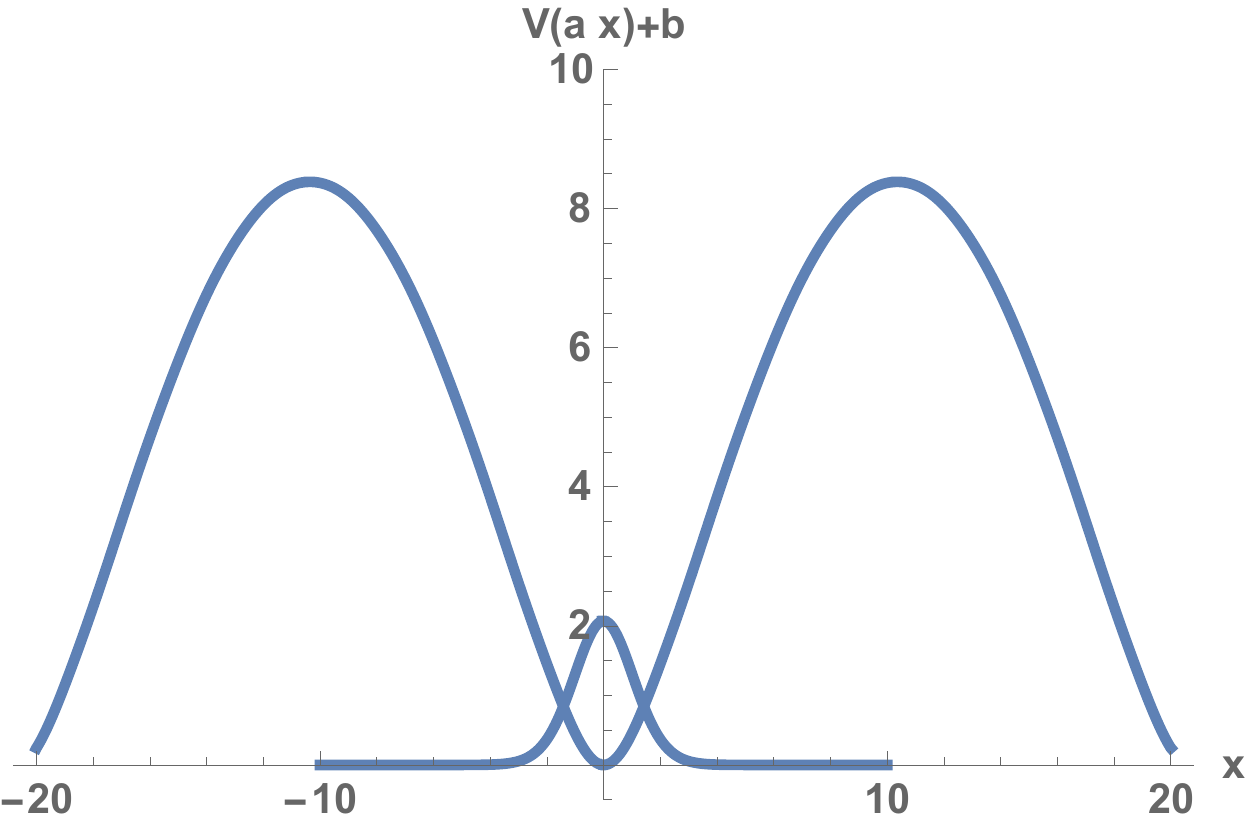}
    \end{minipage}
    \caption{Potential at finite temperature for $SU(2)$ Effective Matrix Model with one fermion in the fundamental representation.The ground state wave function is localized in the valleys  of the potential.}
    \label{3-qubit_unit}
\end{figure}
\begin{figure}
    \centering
    \begin{minipage}[b]{0.7\linewidth}
      \centering
      \includegraphics[width=\linewidth]{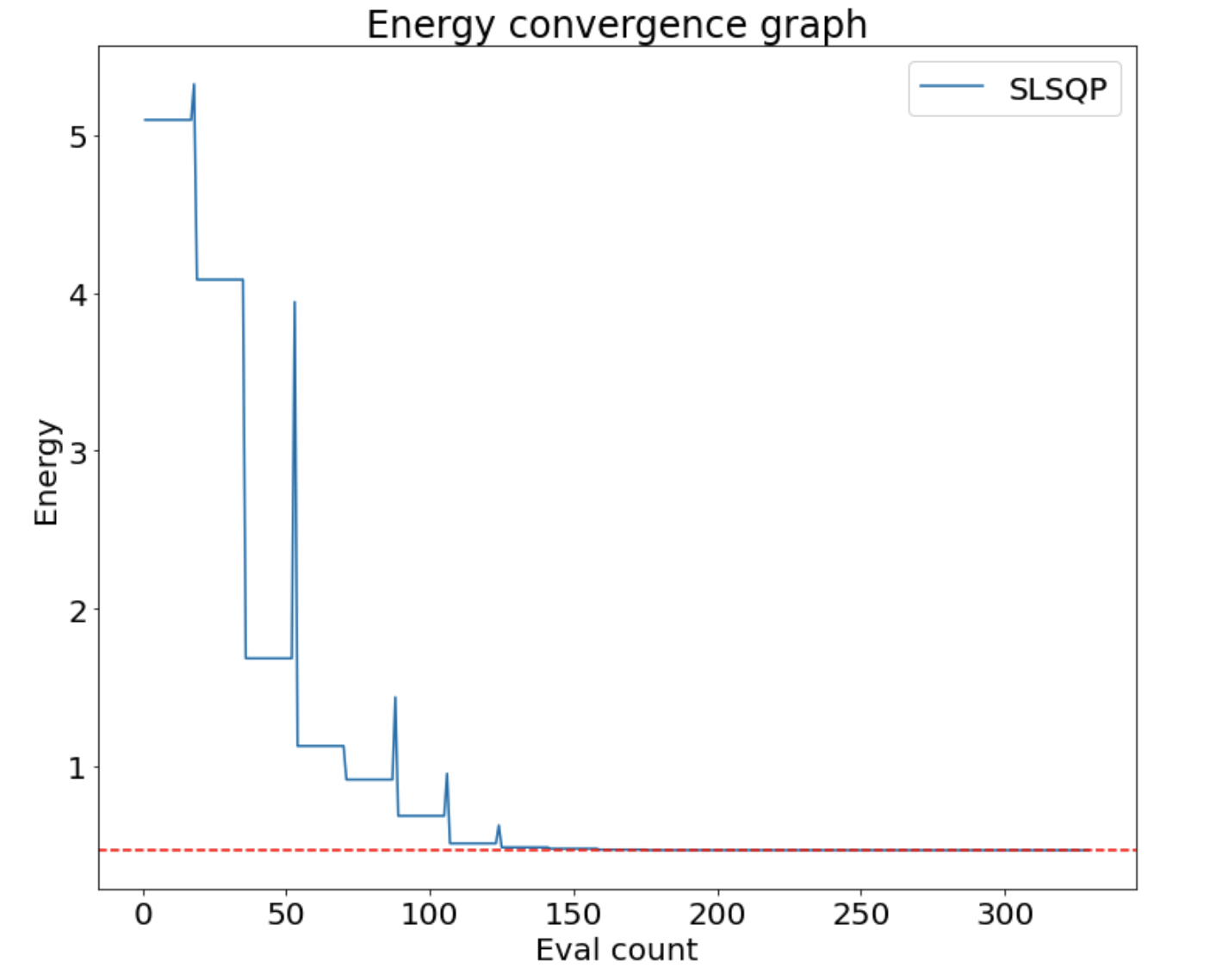}
    \end{minipage}
    \caption{Convergence graph for $SU(2)$ Effective Matrix Model at finite temperature with one fermion in the fundamental representation.}
    \label{3-qubit_unit}
\end{figure}
\begin{table}[ht]
\centering
\begin{tabular}{|l|l|l|l|l|}
\hline
Hamiltonian       & Qubits & Paulis  & Exact Result & VQE Result \\ \hline
$SU(2)$ finite temperature  & 4 & 55 & .46617183 & 0.46617228\\ 
 \hline

\end{tabular}
\caption{\label{tab:BasisCompare}  VQE results for effective Matrix model for $SU(2)$ gauge theory with a fermion in the fundamental representation at finite temperature and using the oscillator basis. The Hamiltonian were mapped to 4-qubit operators with 55 Pauli terms.  The quantum circuit for each simulation utilized an \(R_y\) variational form, with a fully entangled circuit of depth 3. The backend used was a statevector simulator. The Sequential Least SQuares Programming (SLSQP) optimizer was used, with a maximum of 600 iterations.}
\end{table}

\section{VQE for effective Matrix Model with finite density}

One can derive the one-loop Effective Matrix Model potential for finite density by using the imaginary time formulation and introducing an complex vector potential in the imaginary time direction so the covariant derivative is modified by${D_0} = {\partial _0} - i{A_0} $ with ${A_0} =  - i\mu $. The fermion effective Matrix potential at finite temperature and density can then be expressed as \cite{Shiraishi:1986wu}:
\begin{align}
&{V_f}(\phi ,T \ne 0,\mu  \ne 0) = \nonumber\\
&{\mathop{\rm Re}\nolimits} \left[ {\frac{{4{N_F}}}{\beta }\frac{V}{{{{(2\pi )}^3}}}\int_0^\infty  {\frac{{dt}}{t}} \int {{d^3}k\sum\limits_{m,\ell  -  - \infty }^\infty  {\exp \left\{ { - t\left( {{{k^2+\left( {  \frac{{(2n + 1)\pi }}{\beta } + i\mu } \right)}^2} + {{\left( {\frac{{2\pi \ell }}{L} + \frac{\phi }{L}} \right)}^2}} \right)} \right\}} } } \right]\nonumber\\
\end{align}
For $T=0$ and $\mu \ne 0$ one has the simplified form:
\begin{align}
& {V_f}(\phi ,T = 0,\mu  \ne 0) = \nonumber\\
&{N_F}V L\frac{1}{{{L^4}{\pi ^2}}}\left[ {\frac{1}{6}\left\{ {2{\pi ^2}{{\left( {\phi  - \pi } \right)}^2} - {{\left( {\phi  - \pi } \right)}^4} - \frac{7}{{15}}{\pi ^4}} \right\} - \frac{\pi }{3}{{\left( {\phi  - \mu L} \right)}^2}(2\phi  + \mu L)} \right], 0<\phi \leq \mu L\nonumber\\
&{N_F}V L\frac{1}{{{L^4}{\pi ^2}}}\left[ {\frac{1}{6}\left\{ {2{\pi ^2}{{\left( {\phi  - \pi } \right)}^2} - {{\left( {\phi  - \pi } \right)}^4} - \frac{7}{{15}}{\pi ^4}} \right\}} \right], \mu L < \phi \leq 2\pi - \mu L\nonumber\\
&{N_F}VL\frac{1}{{{L^4}{\pi ^2}}}\left[ {\frac{1}{6}\left\{ {2{\pi ^2}{{\left( {\phi  - \pi } \right)}^2} - {{\left( {\phi  - \pi } \right)}^4} - \frac{7}{{15}}{\pi ^4}} \right\} - \frac{\pi }{3}{{\left( {2\pi  - \phi  - \mu L} \right)}^2}(4\pi  - 2\phi  + \mu L)} \right], 2\pi - \mu L < \phi \leq 2 \pi\nonumber\\
\end{align}
We plot this potential in figure 6.
\begin{figure}
    \centering
    \begin{minipage}[b]{0.7\linewidth}
      \centering
      \includegraphics[width=\linewidth]{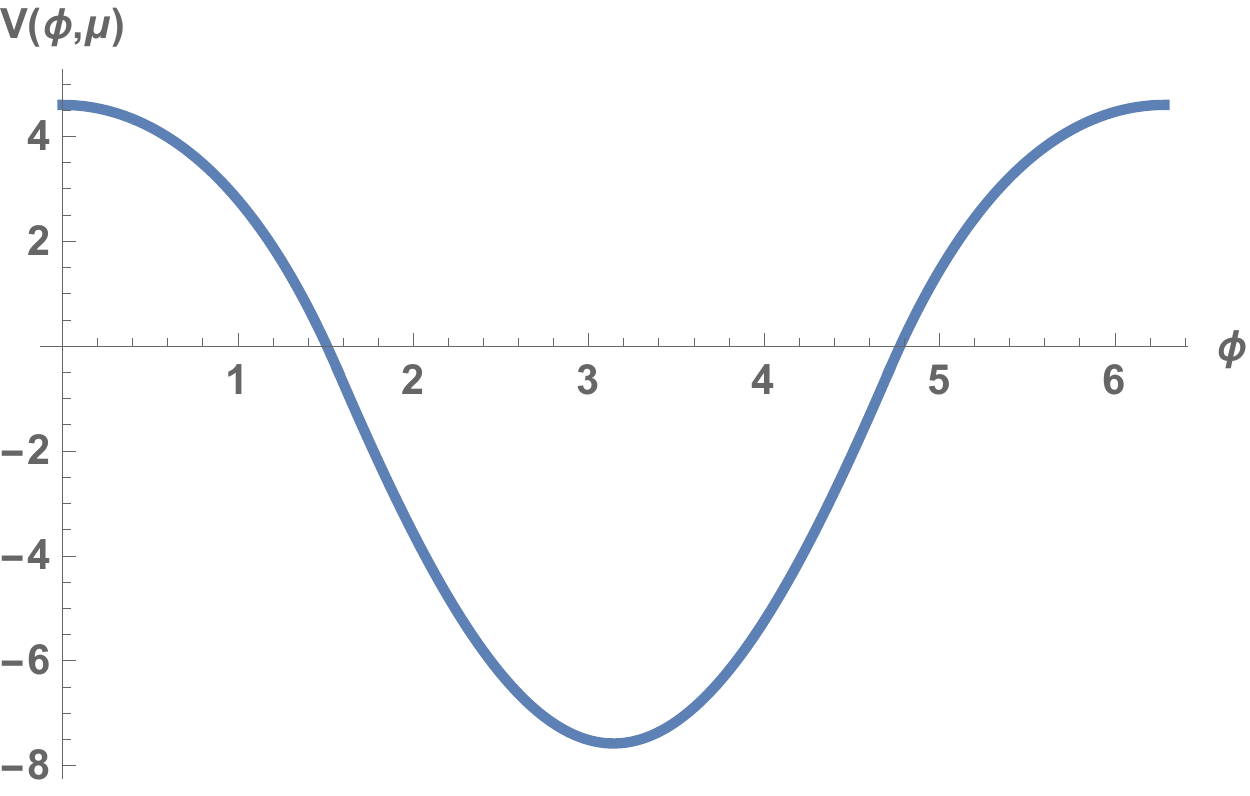}
    \end{minipage}
    \caption{Effective Matrix potential at finite density for $SU(2)$ Effective Matrix Model with one fermion in the fundamental representation.}
    \label{3-qubit_unit}
\end{figure}
\begin{figure}
    \centering
    \begin{minipage}[b]{0.7\linewidth}
      \centering
      \includegraphics[width=\linewidth]{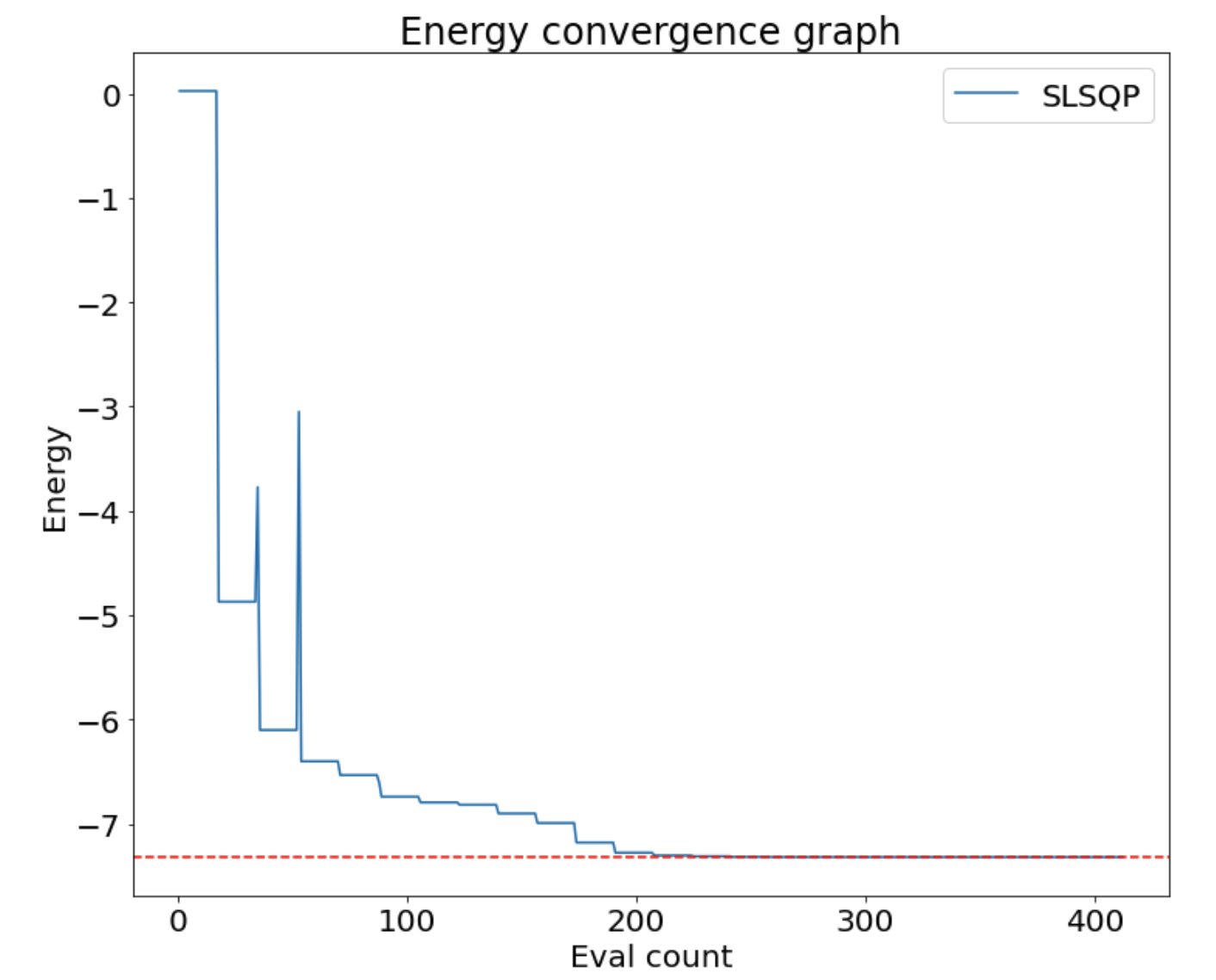}
    \end{minipage}
    \caption{Covergence graph  for $SU(2)$ Effective Matrix Model at finite density with one fermion in the fundamental representation.}
    \label{3-qubit_unit}
\end{figure}
The exact ground state energy $E_0(\mu = \frac{\pi}{2}) = -7.320518$. Using the VQE algorithm we find the results from table 3 which are  in excellent agreement with the classical computation.
\begin{table}[ht]
\centering
\begin{tabular}{|l|l|l|l|l|}
\hline
Hamiltonian       & Qubits & Paulis  & Exact Result & VQE Result \\ \hline
$SU(2)$ finite density  & 4 & 55 & -7.32051788 & -7.32051782\\ 
 \hline

\end{tabular}
\caption{\label{tab:BasisCompare}  VQE results for effective Matrix model for $SU(2)$ gauge theory with a fermion in the fundamental representation at finite density and using the oscillator basis. The Hamiltonian were mapped to 4-qubit operators with 55 Pauli terms.  The quantum circuit for each simulation utilized an \(R_y\) variational form, with a fully entangled circuit of depth 3. The backend used was a statevector simulator. The Sequential Least SQuares Programming (SLSQP) optimizer was used, with a maximum of 600 iterations.}
\end{table}

\section{ VQE for effective Matrix model for SU(3) } 

\begin{figure}[!htb]
\centering
\minipage{0.5\textwidth}
  \includegraphics[width=.8\linewidth]{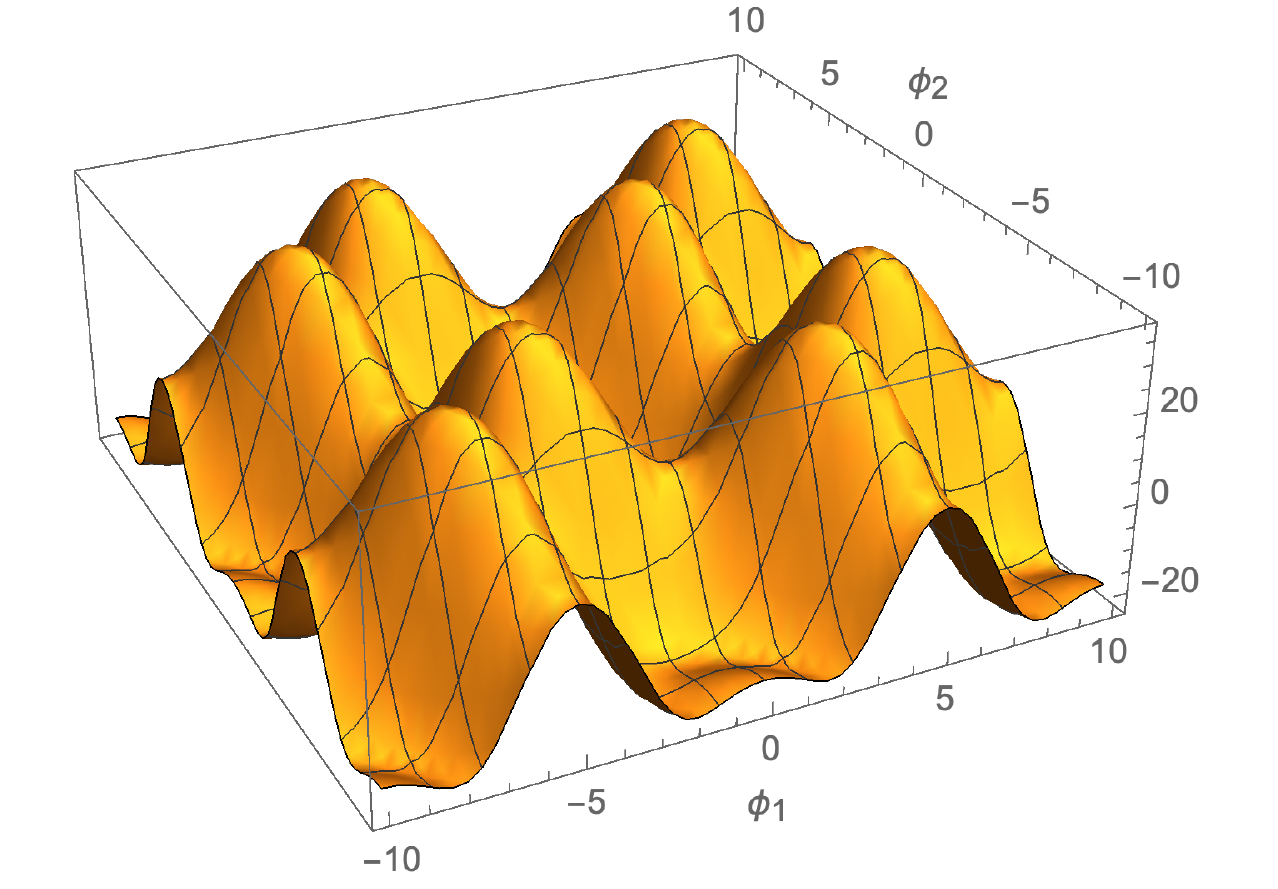}
\endminipage\hfill
\minipage{0.5\textwidth}
  \includegraphics[width=.6\linewidth]{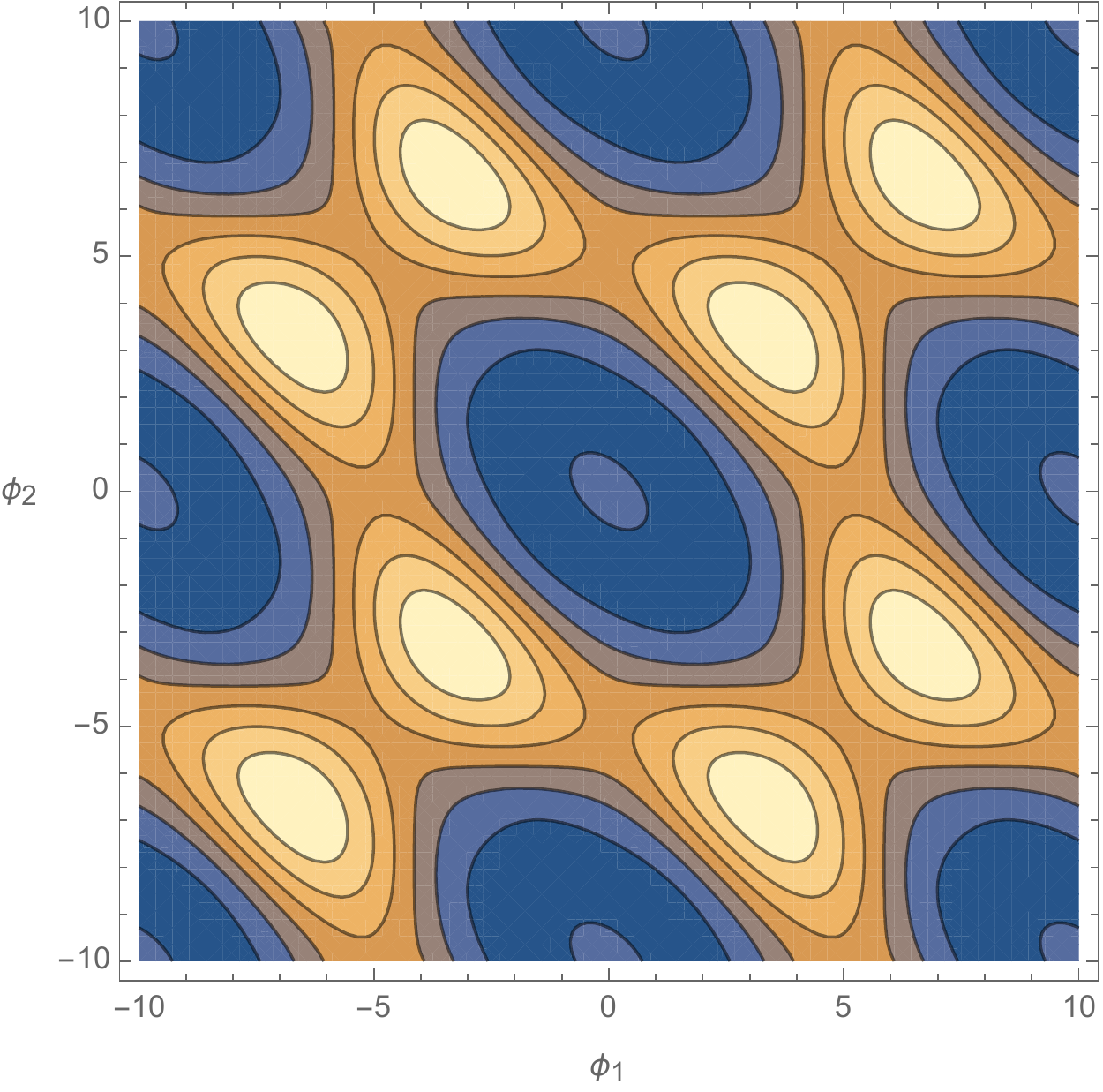}
\endminipage\hfill
\minipage{0.5\textwidth}
  \includegraphics[width=1\linewidth]{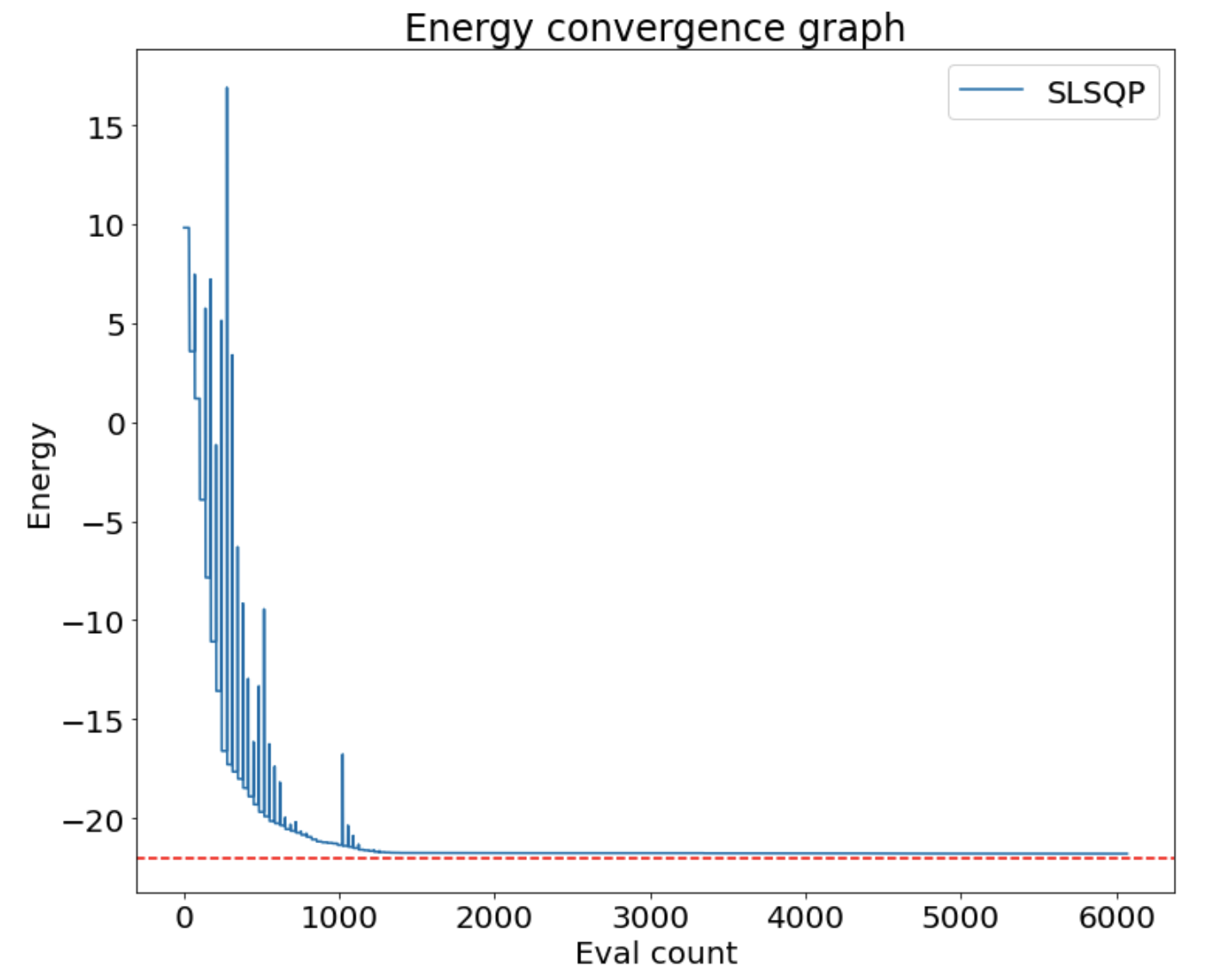}
\endminipage\hfill
\minipage{0.5\textwidth}
  \includegraphics[width=1\linewidth, angle=270]{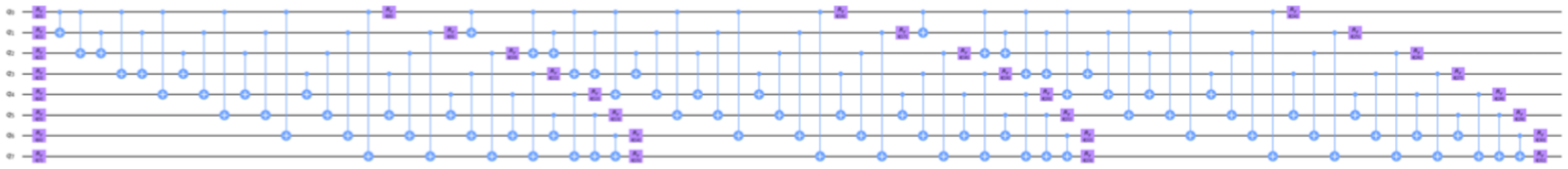}
\endminipage\hfill
\caption{(Upper left) 3D plot of the potential for the effective matrix model for $SU(3)$. (Upper right) Contour plot of the potential for the effective matrix model for $SU(3)$. (Lower left) Convergence plot for $SU(3)$ effective matrix model. (Lower right) Quantum circuit representing the variational ground state for the $SU(3)$ effective matrix model.}
\end{figure}
\begin{table}[ht]
\centering
\begin{tabular}{|l|l|l|l|l|}
\hline
Hamiltonian       & Qubits & Paulis  & Exact Result & VQE Result \\ \hline
$SU(3)$ with fermion in fundamental  & 8 & 9137 & -21.98808168 & -21.793084965\\ 
 \hline
\end{tabular}
\caption{\label{tab:BasisCompare}  VQE results for effective Matrix model for $SU(3)$ gauge theory with a fermion in the fundamental representation using the oscillator basis. The Hamiltonian were mapped to 8-qubit operators with 9137 Pauli terms.  The quantum circuit for each simulation utilized an \(R_y\) variational form, with a fully entangled circuit of depth 3. The backend used was a statevector simulator. The Sequential Least SQuares Programming (SLSQP) optimizer was used, with a maximum of 600 iterations.}
\end{table}
For $SU(3)$ one can proceed similarly to $SU(2)$ except $SU(3)$ is rank two and there are two Wilson lines $\phi_1$ and $\phi_2$. The $SU(3)$ gauge potential for the Effective Matrix model is parameterized as:
\begin{equation}A_3^a{\lambda ^a} = \left( {\begin{array}{*{20}{c}}
{{\phi _1}}&0&0\\
0&{{\phi _2}}&0\\
0&0&{ - {\phi _1} - {\phi _2}}
\end{array}} \right)\end{equation}
and Effective Matrix potential for $SU(3)$ with one fermion in the fundamental representation is:
\begin{equation}{V_{eff}}({\phi _1},{\phi _2}) = V_{eff}^g({\phi _1},{\phi _2}) + V_{eff}^f({\phi _1},{\phi _2})\end{equation}
where:
\[V_{eff}^g({\phi _1},{\phi _2}) =  - \frac{2}{{{\pi ^2}}}\frac{1}{{{L^4}}}\sum\limits_{j,k}^3 {\sum\limits_{m = 1}^\infty  {\frac{{\cos ({\phi _j} - {\phi _k})}}{{{m^4}}}} } \]
\begin{equation}V_{eff}^f({\phi _1},{\phi _2}) =  - \frac{2}{{{\pi ^2}}}\frac{1}{{{L^4}}}\sum\limits_j^3 {\sum\limits_{m = 1}^\infty  {\frac{{\cos ({\phi _j})}}{{{m^4}}}} } \end{equation}
and ${\phi _3} =  - {\phi _1} - {\phi _2}$. Because there are two Wilson line variables we need to use tensor products to construct the Hamiltonian for the VQE. Defining:
\[{\phi _1} = {\phi} \otimes I\]
\[{\phi _2} = I \otimes {\phi}\]
\[{p_{{\phi _1}}} = {p_{\phi}} \otimes I\]
\begin{equation}{p_{{\phi _2}}} = I \otimes {p_{\phi}}\end{equation}
where $\phi$  and $p_{\phi}$ are given by 2.7 amd 2.8. The Hamiltonian is then written  as
\begin{equation}{H_{SU3}} = \frac{1}{2}p_{{\phi _1}}^2 + \frac{1}{2}p_{{\phi _2}}^2 + {V_{eff}}({\phi _1},{\phi _2})\end{equation}
Using $16\times 16$ matrices for $\phi_1$ and $\phi_2$ so the total Hamiltonian is $256 \times 256$ and can be represented by eight qubits. Using the VQE and the SLSQP optimizer we find the results in table 4 which are  in excellent agreement with classical computation.

\newpage

\section{Conclusions}

In this paper we studied the Effective Matrix Model for gauge theories on a quantum computer. We were able to obtain accurate results for $SU(2)$ and $SU(3)$ gauge models including fermion including finite temperature and finite density effects involving nonzero chemical potential using the Variational Quantum Eigensolver (VQE) approach on a IBM quantum computer. It will be interesting to extend computations to include new terms in the Matrix model coming from nonperturbative effects in the equation of state as was done in \cite{Dumitru:2012fw}. Finally by considering inhomogeneous Matrix models which depend on one spatial coordinate, or by studying large rank groups as occur in UV complete theories like string theory or in strongly interacting dark matter models, the effective matrix model may exceed the simulation capabilities of classical computers and thus provide an excellent  opportunity for quantum advantage on quantum computers.

\section*{Acknowledgements}
This material is based upon work supported by the U.S. Department of Energy, Office of Science, National Quantum Information Science Research Centers, Co-design Center for Quantum Advantage (C2QA) under contract number DE-SC0012704. This project was supported in part by the U.S. Department of Energy, Office of Science, Office of Workforce Development for Teachers and Scientists (WDTS). We wish to acknowledge useful discussions on effective Matrix models with Rob Pisarski.

\end{document}